%% file: conference_101719.tex
\documentclass[conference]{IEEEtran}
\IEEEoverridecommandlockouts
\usepackage{cite}
\usepackage{amsmath,amssymb,amsfonts}
\usepackage{algorithmic}
\usepackage{graphicx}
\usepackage{textcomp}
\usepackage{xcolor}
\usepackage{acro}
\usepackage{url}
\usepackage{multirow}
\usepackage{titlesec}
\usepackage{enumitem}
\usepackage{soul}
\usepackage{amsmath}
\usepackage{stfloats}
\usepackage[caption=false, font=footnotesize]{subfig}

\def\BibTeX{{\rm B\kern-.05em{\sc i\kern-.025em b}\kern-.08em
    T\kern-.1667em\lower.7ex\hbox{E}\kern-.125emX}}

\input{acronyms}

\begin{document}

\setlength{\abovedisplayskip}{5pt}
\setlength{\belowdisplayskip}{5pt}
\titlespacing{\section}{0pt}{*1.0}{*1}
\titlespacing{\subsection}{0pt}{*0.8}{*0.6}

\title{En-route Charging Coordination for Electric Trucks}

\author{\IEEEauthorblockN{Joas Kahlert}
\IEEEauthorblockA{\textit{Decision and Control Systems}\\
\textit{KTH Royal Institute of Technology}\\
Stockholm, Sweden \\
kahlert@kth.se}
\and
\IEEEauthorblockN{Ruiting Wang}
\IEEEauthorblockA{\textit{Decision and Control Systems}\\
\textit{KTH Royal Institute of Technology}\\
Stockholm, Sweden \\
ruiting@kth.se}
\and
\IEEEauthorblockN{Jonas Mårtensson}
\IEEEauthorblockA{\textit{Decision and Control Systems}\\
\textit{KTH Royal Institute of Technology}\\
Stockholm, Sweden \\
jonas1@kth.se}
}

\maketitle

\begin{abstract}
The electrification of long-haul freight transport introduces several new challenges, such as the limited capacity and congestion at en-route charging infrastructure. To reduce waiting times during peak periods, this paper proposes a framework for coordinated charging scheduling. The approach employs a mixed-integer formulation to optimize charging-related costs across charging, operation, battery degradation, and congestion delay, considering a range of scenarios. The results demonstrate that coordinated scheduling yields substantial cost savings up to 36\% compared to uncoordinated scheduling, particularly by reducing battery degradation and delay costs. 

\end{abstract}

\begin{IEEEkeywords}
Electric trucks, charging coordination, charging scheduling, mixed integer programming, optimization 
\end{IEEEkeywords}

\section{Introduction}

The electrification of the heavy-duty transport sector has gained significant momentum in response to the growing demand for sustainable mobility solutions \cite{iea_2025}. This transition is driven not only by regulatory pressures, such as EU directives and national transport policies, but also by evolving customer preferences and rising public expectations \cite{iea_2025}. Electrification is therefore no longer merely optional for freight carriers; it is becoming essential for maintaining competitiveness and long-term viability. However, freight companies that aim to electrify their operations face substantial financial risks when investing in these emerging technologies prematurely \cite{bjorklund_2025}. The large-scale rollout of electrified transport solutions is hindered by several practical challenges, such as high upfront investments, long charging times, and limited driving ranges \cite{bjorklund_2025}. Unlike their fossil-fuel counterparts, \acp{BET} require more careful operational planning to ensure that the additional charging requirements are effectively integrated into their transport schedules. Moreover, long-haul operations depend on en-route charging stations with limited capacity that operate in a highly competitive environment.  

Due to the added complexity of these charging decisions, it is no longer suitable for the decisions to be made by the truck driver alone. This creates the need for a scheduling system capable of computing and continuously providing charging instructions to drivers. Such scheduling can become highly complex, as it depends on a wide range of operational conditions related to both trucks and charging stations. Problems of this domain can broadly be categorized as \acp{E-VSP}.

One common approach is to optimize each truck’s schedule independently by using historical demand patterns. Authors in \cite{bai_2025} propose a decentralized method where trucks request estimated waiting times based on predicted arrivals, while \cite{barba_2025} focuses on incorporating dynamic energy prices \ac{HoS} regulations. Because \ac{HoS} rules require drivers to take periodic rest breaks during long trips, it is often most time-efficient to align truck charging schedules with these mandatory rest periods. Although such strategies avoid assuming that trucks share their charging intentions, decentralized approaches remain inherently limited in their ability to actively mitigate peak loads in real time. Instead, they can only plan around anticipated congestion based on forecasts, rather than dynamically coordinating charging decisions to alleviate emerging bottlenecks.

In contrast, studies such as \cite{alam_2023, alhanahi_2024} adopt centralized \ac{E-VSP} frameworks with jointly optimized schedules for multiple trucks. Authors in \cite{alam_2023} optimize charging and platooning strategies for long-haul \acp{BET}, while \cite{alhanahi_2024} propose a strategy integrating both public and private charging for large \ac{BET} fleets. However, these works focus on broad operational and logistical dimensions and do not fully address key aspects of en-route fast charging, such as limited station capacities and congestion dynamics. As en-route charging becomes increasingly important with the electrification of long-haul freight transport, there is a clear need for coordinated charging at en-route stations to efficiently manage rising demand. To address this gap, this study specifically:
\begin{itemize}
    \item Formulate the \ac{E-VSP} for a sequence of en-route charging stations as an \ac{MIP}. 
    \item Explicitly incorporate charging constraints, \ac{HoS} requirements, queue dynamics, time windows, and battery degradation in the model formulation. 
    \item Evaluate the coordinated approach against a decentralized reference strategy across multiple charging scenarios, and quantify its cost-saving potential.  
\end{itemize}

%Recently, studies employing reinforcement learning for smart charging strategies have also demonstrated notable improvements \cite{tuchnitz_2021, sultanuddin_2023}. Although highly effective in handling the stochastic operational environment at charging sites, these approaches often struggle to provide the required level of detail and can be difficult to interpret.

\section{Problem Description}

To frame the \ac{E-VSP}, we consider a pre-established charging corridor, that is, a sequence of charging stations along a route. For example, this may correspond to a longer section of highway. \acp{BET} traverse this corridor in both directions and must periodically recharge at these stations to maintain a sufficient \ac{SoC}, while also taking rest breaks in accordance with \ac{HoS} regulations. If a truck arrives at a charging station with no available chargers, it must wait until one becomes available.

The objective is to efficiently coordinate the charging schedules of all trucks while minimizing their operational costs, including delay penalties, charging expenses, and battery degradation costs. To achieve this, we consider a centralized scenario in which trucks traversing the corridor communicate their charging requirements to a cloud-based controller. The controller can automatically co-optimize and broadcast individualized charging and rest instructions upon request.

For the decentralized reference scenario, we assume that trucks operate solely in their own interest and do not share their charging intentions. Under this assumption, each truck independently recharges whenever its remaining energy is insufficient to reach the next charging station along its path. The truck then rests and charges only as much as necessary to reach its exit point from the corridor. As discussed in the next section, the centralized strategy also incorporates a term that minimizes the required amount of charged energy indirectly. As a result, the total energy charged under both strategies remains the same, with the only difference being the timing and location of the charging.

\subsection{Mathematical Formulation}

We model the problem as an \ac{MIP}. Many of the modeling components introduced below follow conventions that have become standard in the literature, and we therefore omit nonessential details. For notation, see Table \ref{tab:notation}.

\renewcommand{\arraystretch}{0.92} % tighter rows

\vspace{-6pt}
\begin{table}[htbp]
\centering
\caption{Notation}
\label{tab:notation}
\vspace{-6pt}
\begin{tabular}{c|p{2.2cm}||c|p{2.6cm}}
\hline
\textbf{Vars.} & Description & \textbf{Params.} & Description \\
\hline
$x_{n,i}$ & Charging indicator
& $d_{i,j}$ & Segment distance \\

$y_{n,i}$ & Resting indicator
& $\kappa$ & \ac{SoC} safety factor \\

$z_{n,i}$ & Visit indicator
& $\hat E_n$& Battery capacity\\

$\chi_{n,i,k}$ & Charger indicator
& $\Delta E_n$ & Energy consumption\\

$\omega_n$ & Delay indicator
& $v_n$ & Truck speed \\

$E_{n,i}$ & Remaining energy
&&\\

$\Delta e_{n,i}$ & Energy charged
& $h$ & Rest time \\

$\tau_{n,i}$ & Charging time
& $\overline W$ & Exit time window \\

$\zeta_{n,i}$ & Charger occupation time
& $\delta^1_i$/$\delta^2_i$ & Charging/visit overhead \\

$\overline t_{n,i}/\underline t_{n,i}$ & Arrival/departure time
& $\rho_n^{opr}$& Operating cost per time unit \\

$t^*_{n,i}$ & Charging start
& $\rho_n^{dly}$& Delay penalty\\

$\overline \xi_{n,i}/\underline \xi_{n,i}$ & \ac{SoC} before/after charging
& $\rho_i^{lct}$ & Electricity cost per time unit\\

$\overline C_{n,i}/\underline C_{n,i}$ & Cost of \ac{SoC} state before/after charging
& $\hat P_n/\tilde P_i$ & Max accepted/supplied charging power  \\

\hline
\end{tabular}
\end{table}
\vspace{-6pt}

We consider a set of $n\in\mathcal N$ \acp{BET} that traverse a charging corridor with the ordered set of charging stations $i\in\mathcal I$. Because trucks may travel in different directions along the corridor, truck $n$ encounters the stations in a direction-dependent order $\mathcal{I}_n$. Station $i^0$ and $i^e$ are the first and last stations on the corridor, while by the indexation $i^-/i^+$ we mean the previous or next station in order of station $i$. The set $\mathcal{I}^-_n = \mathcal{I}_n \setminus \{i^e\}$ contains all stations except the last one on the corridor.  

With this definition, the \textbf{battery \ac{SoC} dynamics} are,
\begin{subequations}
    \label{eq:charge}
    \begin{align}
        & E_{n,i^+}\!=\!E_{n,i}\!-d_{i,i^+}\Delta E_{n}\!+\!\Delta e_{n,i},\!\! &\forall (n,i)\in \mathcal N\times\mathcal I_n^-,\label{eq:char_event}\\
        & \kappa \hat E_n\leq E_{n,i}\leq \hat E_n, &\forall (n,i)\in \mathcal N\times\mathcal I_n,\label{eq:SoC}\\  
        &\Delta e_{n,i} = \tau_{n,i}\min\{\hat P_n, \tilde P_{i}\},&\forall (n,i)\in \mathcal N\times\mathcal I_n,\label{eq:energy_bound_1}\\
        & \Delta e_{n,i}\leq x_{n,i}\hat E_n,&\forall (n,i)\in \mathcal N\times\mathcal I_n,\label{eq:energy_bound_2}\\
        &  0\leq\Delta e_{n,i}\leq \hat E_n-E_{n,i},&\forall (n,i)\in \mathcal N\times\mathcal I_n.\label{eq:energy_bound_3}
    \end{align}
\end{subequations}

Eq.~\eqref{eq:char_event} represent charging and discharging events, Eq.~\eqref{eq:SoC} expresses \ac{SoC} limits, and Eq.~\eqref{eq:energy_bound_1}-\eqref{eq:energy_bound_3} binds charging energy. The supplied power is assumed to be constant. 

The \textbf{time dynamics} for driving, charging, and resting (e.g. \ac{HoS} constraint) are given by,
\begin{subequations}
    \label{eq:times}
    \begin{align}
        &\overline t_{n,i^+}= \underline  t_{n,i}+\frac{d_{i,i^+}}{v_n},&\forall (n,i)\in \mathcal N\times\mathcal I_m^-,\label{eq:next_station}\\
        & \underline t_{n,i}= t^*_{n,i}+\zeta_{n,i}+\delta^1_i x_{n,i},&\forall (n,i)\in \mathcal N\times\mathcal I_n,\label{eq:visit_time_1}\\
        &t^*_{n,i} \geq  \overline t_{n,i} + \delta^2_iz_{n,i},&\forall (n,i)\in \mathcal N\times\mathcal I_n,\label{eq:visit_time_2}\\
        &\zeta_{n,i}\geq\tau_{n,i},&\forall (n,i)\in \mathcal N\times\mathcal I_n,\label{eq:zeta_1}\\
        &\zeta_{n,i}\geq hy_{n,i},&\forall (n,i)\in \mathcal N\times\mathcal I_n,\label{eq:zeta_2}\\
        &\sum_{i\in\mathcal I_{t}}y_{n,i}\geq 1,&\forall n\in\mathcal N,\label{eq:must_rest}\\
        &z_{n,i}\geq x_{n,i},\ y_{n,i},&\forall (n,i)\in \mathcal N\times\mathcal I_n,\label{eq:if_visit}
    \end{align}
\end{subequations}

Eq.~\eqref{eq:next_station}-\eqref{eq:visit_time_2} model the full station-to-station time relations. The charger occupation time is $\zeta_{n,i}=\max(\tau_{n,i}, h)$ (Eq.~\eqref{eq:zeta_1}-\eqref{eq:zeta_2}). $z_{n,i}$ indicates if the station is visited for charging $x_{n,i}$ or for resting $y_{n,i}$ (Eq.~\eqref{eq:if_visit}). We also assume that each driver must take at least one rest break of length $h$ (Eq.~\eqref{eq:must_rest}).

We model the limited charging station capacity by prohibiting charging time windows $[t^*_{n,i}, \underline t_{n,i}]$ to overlap on the same charger for any two trucks $n$ and $m$. In this way, trucks must either wait in the time window  $[\overline t_{n,i},  t^*_{n,i}]$ until a time slot becomes available, or charge at a different charger or station entirely. This relation can be expressed by,

\begin{subequations}
    \label{eq:nonoverlap}
    \begin{align}
    \begin{split}
        &(\underline t_{n,i} \leq t^*_{m,i})\vee(\underline t_{m,i} \leq t^*_{n,i})\quad\text{if}\quad\chi_{n,i,k}\wedge\chi_{m,i,k}\\
        & \qquad\qquad\quad\ \ \forall ((n,m),i,k)\in\mathcal N\times\mathcal I_n\times\mathcal K_i,\ n \neq m,\\
    \end{split}\label{eq:overlap}\\
    &x_{n,i}=\sum_{k\in\mathcal K_i}\chi_{n,i,k},\qquad\qquad\qquad\quad\, \forall (n,i)\in \mathcal N\times\mathcal I_n,\label{eq:station_charger}
    \end{align}
\end{subequations}

which can conventionally be linearized through the introduction of properly chosen auxiliary variables and big-M notation. Notably, Eq. \eqref{eq:overlap} does not enforce any specific charging order but only reserves one charging slot per charger for any given time. Eq. \eqref{eq:station_charger} makes sure the charging station and station charger indicators match for any given charging event. Furthermore, if a truck is not able to make the expected upper time window, then we enforce a penalty through
\begin{equation}
    \label{eq:delay}
    w_n=
    \begin{cases}
        1,\quad \text{if}\quad t_{n,i^e}>\overline W,\\
        0,\quad \text{if}\quad t_{n,i^e}\leq\overline W,\\
    \end{cases}
    \quad \forall n\in \mathcal N.
\end{equation}

To incorporate battery degradation costs for each charging event, we adopt the formulation proposed in \cite{lee_2022}. The authors introduce an alternative to the computationally intensive rainflow counting method by deriving a one-cycle battery cost function $C^B$ based on the cycle life curve $N^C$ and a cleverly constructed auxiliary \ac{SoC} state. For further details, we refer the reader to \cite{lee_2022}.
\begin{equation}
    \label{eq:CB}
    \begin{aligned}
        C^B(\overline \xi,\underline \xi)&=
        \begin{cases}
            C(\overline\xi)-C(\underline \xi),&\text{charge event},\\
            0, & \text{discharge event},\\
        \end{cases}\\
        N^C(\xi)&=\beta_0\xi^{-\beta_1}e^{\beta_2(1-\xi)},\quad C(\xi)=\frac{C^{cap}}{N^{C}(\xi)}.\\
    \end{aligned}
\end{equation}

Here, $\overline \xi$, $\underline\xi$ represents the \ac{SoC} before, and after charging. Since the battery state in \eqref{eq:charge} is updated only once per station visit $i$, we must decompose the charging events into separate before- and after charge states in order to align with the degradation cost formulation given in \eqref{eq:CB}.
\begin{subequations} \label{eq:batterydeg}
    \begin{align}
        &\overline \xi_{n,i}=\frac{E_{n,i}}{\hat E_n},&\forall (n,i)\in \mathcal N\times\mathcal I_n,\\
        &\underline \xi_{n,i}=\frac{E_{n,i}+\Delta e_{n,i}}{\hat E_n},&\forall (n,i)\in \mathcal N\times\mathcal I_n,\\
        &\overline C_{n,i}=\mathcal P \mathcal W\mathcal L_R(C;\overline \xi_{n,i}),&\forall (n,i)\in \mathcal N\times\mathcal I_n,\\
        &\underline C_{n,i}=\mathcal P \mathcal W\mathcal L_R(C;\underline \xi_{n,i}),&\forall (n,i)\in \mathcal N\times\mathcal I_n.
    \end{align}
\end{subequations}
where for convenience of notation we define $\mathcal P \mathcal W\mathcal L_{R}[f;x]$ as the piecewise linear interpolant of $f$ with $R$ breakpoints. Such interpolants can routinely be implemented in \acp{MIP} through standard SOS2-based constraints.

For completeness, the variable domain constraints are,
\begin{equation}
    \label{eq:domains}
    \begin{aligned}
        &E_{n,i},\ \Delta e_{n,i},\ \tau_{n,i},\ \zeta_{n,i},\ \underline t_{n,i},\ t^*_{n,i},\ \overline t_{n,i},\ \overline C_{n,i},\ \underline C_{n,i}\in\mathbb R^+,\\
        &\xi_{n,i}\in[0,1],\\
        &x_{n,i},\ y_{n,i},\ z_{n,i},\ \chi_{n,i,k},\ \omega_n\in\{0,1\}.
    \end{aligned}
\end{equation}

The objective is to minimize the combined costs of all trucks 
\begin{equation}
    \label{eq:obj}
    \begin{split}
        J^{obj} =& \sum_{n\in\mathcal N}\Big[\rho_n^{opr}\big(\underline t_{n,i^e}-\overline t_{n,i^0}\big)+\rho_n^{dly}w_n\\
        &\quad+\sum_{i\in\mathcal I_{n}}\big[\rho^{lct}_{i}\Delta e_{n,i}+(\overline C_{n,i}-\underline C_{n,i})\big]\Big],
    \end{split}
\end{equation}

which consists of: the continuous operation costs over the time in the corridor $\rho_n^{opr}(\underline t_{n,i^e}-\overline t_{n,i^0})$; the penalty costs in case of delay, $\rho_n^{dly}w_n$; the electricity costs $\rho^{lct}_{i}\Delta e_{n,i}$; and the battery degradation costs $\overline C_{n,i}-\underline C_{n,i}$. The full \ac{MIP} model,
\begin{equation*}
\begin{aligned}
    \min\quad &J^{obj},\\
    \text{s.t.}\quad &\eqref{eq:charge},\ \eqref{eq:times},\ \eqref{eq:nonoverlap},\ \eqref{eq:delay},\ \eqref{eq:batterydeg},\ \eqref{eq:domains},\\
\end{aligned}
\end{equation*}

is solved with Gurobi LLC \cite{gurobi}.

\subsection{Scenario Description}

Trucks incoming to the corridor are randomly generated with a time they enter the corridor $\overline t_{n,i^0}\sim U(0,12)$ h and a battery state $E_{n,i^0}\sim U(0.25\hat E_n, \hat E_n)$ kWh. Trucks drive at speed $v_n=85$ km/h, charge at rate $\hat P_n,\tilde P_i = 750$ kW, with battery capacity $\hat E_n=600$ kWh, and consume on average $1.8$ kWh/km. These values are in line with the newest technology and standards \cite{scania_2025}. The safety margin for a lower \ac{SoC} bound is $\kappa = 0.1$, and the rest period is $h=45$ min. Overhead for visiting and charging at a station is $\delta^2_i=7$ min, and $\delta^1_i=5$ min. The electricity cost is $\rho^{lct}_i=0.3$€/kWh, and the continuous operational cost is $\rho^{opr}_n=30$€/h, including e.g. driver salary. The cost for a missed delivery time window can depend on several different factors, like value/type of shipment, contractual terms, and operational factors downstream. For consistency, we assume a fixed penalty, same for all trucks, $\rho_n^{dly}=500$€ \cite{weberlogistics_2025}. We assume that the battery cost is proportional to the battery size with $C^{kWh}= 250$€/kWh. The capital cost of a battery is then $C_n^{cap}=C^{kWh}\hat E_n$. Lastly, the upper delay window $\overline W_n$ is independently calculated based on the total time it would take for each truck to traverse the corridor, in addition to a $25\%$ margin for charging and resting. The charging corridor is $400$ km long and contains five charging stations, each equipped with two chargers, positioned at fixed intervals along the route. In the scenarios analyzed below, we randomly generate and dispatch a set of $|\mathcal N|$ trucks to traverse the corridor. 

\section{Results and Discussion}

\begin{figure}[htbp]
\vspace{-6pt}
    \centering
    \hspace*{-0.2cm}
    \includegraphics[width=1.03\linewidth]{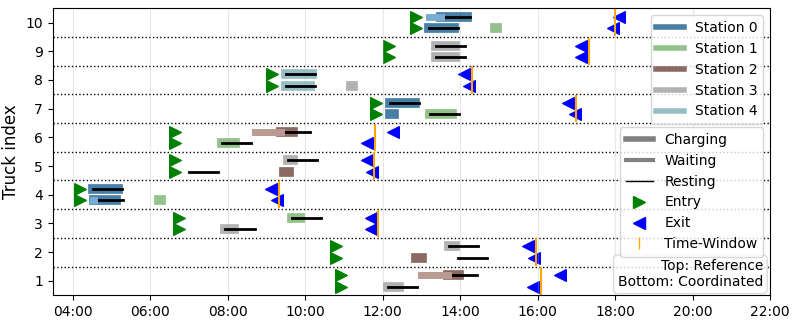}
    \caption{Charging schedules for reference and coordinated scheduling over the studied highway corridor for a 75-truck scenario. Only the schedule of 10 trucks is shown for demonstration purposes. }
    \label{fig:trucks}
\vspace{-6pt}
\end{figure}

By solving the problem under different system configurations, we compare uncoordinated and coordinated scenarios across varying truck volumes along the corridor. For demonstration purposes, the perspective of a subset of individual trucks is visualized in Fig. \ref{fig:trucks}.

As shown in the plot, the longest charging stop generally coincides with the resting period, resulting in significant time savings. Unlike the reference scenario, which by design prioritizes long, uninterrupted charging intervals, the coordinated approach may split charging into separate events. This strategy offers two key advantages: first, it enables more flexible time-slot allocation, preventing individual chargers from being occupied by near hour-long sessions; second, it mitigates battery degradation by maintaining a more uniform \ac{SoC} level, which is less detrimental compared to the deep charging cycles in the reference case. The resulting cost differences are reported in Table~\ref{tab:numerics}.

\begin{table}[htbp]
    \centering
    \vspace{-6pt}
    \caption{Cost Breakdown / Truck [€]}
    \begin{tabular}{c|c|c|c|c|c|c}
        \hline
        Trucks & Method & Charging & Operating & Battery & Delay & Total \\
        \hline
        \multirow{2}{*}{40}
        & ref. & 92 & 150 & 51 & 48 & 341 \\
        & coord. & 92 & 151 & 30 & 0 & 273 \\
        \hline
        \multirow{2}{*}{55}
        & ref. & 94 & 152 & 50 & 95 & 392 \\
        & coord. & 94 & 151 & 31 & 0 & 276 \\
        \hline
        \multirow{2}{*}{70}
        & ref. & 93 & 155 & 51 & 145 & 443 \\
        & coord. & 93 & 151 & 32 & 6 & 283 \\
        \hline
    \end{tabular}
    \label{tab:numerics}
\vspace{-8pt}
\end{table}

If schedules are left uncoordinated, a truck may arrive at a station to find all chargers occupied, forcing it to wait and potentially miss its designated time window at the end of the corridor. An example of this undesired outcome is illustrated by e.g. truck~$\#6$ in Fig.~\ref{fig:trucks}. Missing a time window incurs a penalty cost, the aggregate of which is also reflected in the cost breakdown in Table~\ref{tab:numerics}. Although the reference case exhibits substantial delay-related costs, coordinated schedules result in near-zero delay costs across all scenarios.

\begin{figure*}[t]
    \vspace{-10pt}
    \centering
    \subfloat[Reference/uncoordinated charging schedules]{\includegraphics[height=0.3\linewidth]{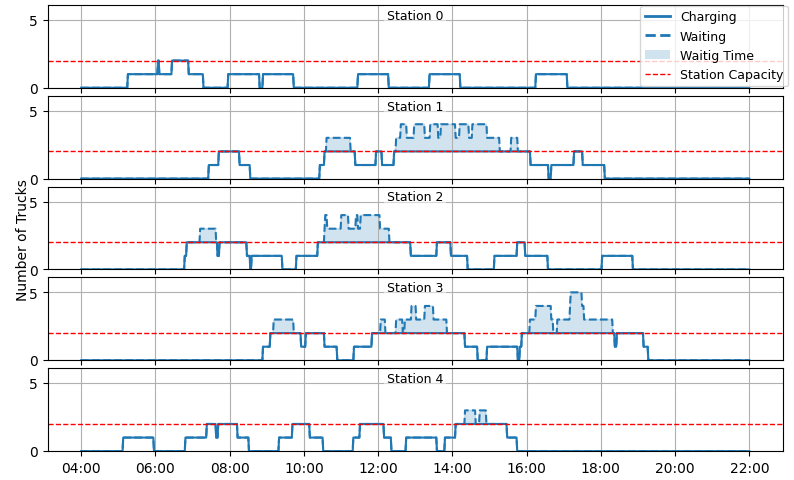}}
    \subfloat[Coordinated charging schedules]
    {\includegraphics[height=0.3\linewidth]{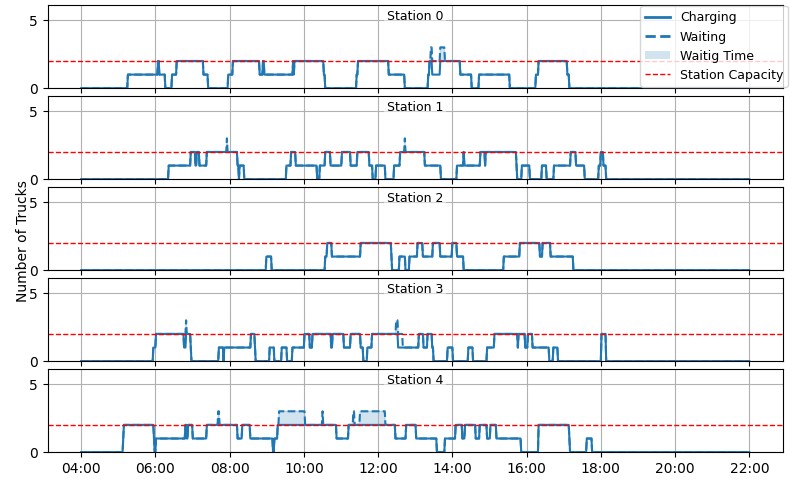}}
    \hfill
    \caption{This figure illustrates the charging station occupancy status over time for a 75-truck scenario. The waiting time shown in the shaded area has reduced significantly by coordinated scheduling.}
    \label{fig:stations}
\vspace{-8pt}
\end{figure*}

The cumulative waiting time caused by queue buildup over station capacity can also be observed from the stations' perspective, as shown in Fig. \ref{fig:stations}. Here, we clearly see that by coordinating charging events, it is possible to schedule all charging requirements in a way that most of the waiting time can be mitigated.

Comparing both methods across scenarios, we observe that coordination becomes increasingly cost-effective as the number of trucks grows. According to our model, we observe total cost savings of $20\%$, $30\%$, and $36\%$ for each scenario, respectively. This difference stems almost exclusively from the battery and delay-related costs as trucks must charge and drivers must be compensated regardless of coordination. Battery degradation costs are reduced by about $37\%$ while delay costs are almost entirely eliminated for all scenarios.

\vspace{-1pt}
\section{Conclusion and Future Work}  
\vspace{-1pt}
This study introduces a framework designed to capture many of the complex operational and logistical considerations part of the \ac{E-VSP}, by considering charging constraints, \ac{HoS} requirements, queue dynamics, time windows, and battery degradation. Results demonstrate that coordinated charging management consistently outperforms decentralized approaches, particularly helpful to mitigate congestion at charging stations when demand increases, and can save up to $36\%$ of costs in comparison to uncoordinated scheduling. 

To extend our work to a more public charging station setting, we are currently working on incorporating the charging requirements of homogeneous trucks at en-route charging stations. In addition, we are investigating semi-centralized coordination schemes that enable internal charging scheduling through less sensitive truck-to-station communication mechanisms.

\vspace{-8pt}

\input{bib.tex}
\vspace{-1pt}

\end{document}

%% file: acronyms.tex
\DeclareAcronym{BET}{
  short = BET ,
  long  = Battery Electric Truck,
  short-plural = s % Plural form of the short form
}

\DeclareAcronym{E-VRP}{
  short = E-VRP ,
  long  = Electric Vehicle Routing Problem,
  short-plural = s % Plural form of the short form
}

\DeclareAcronym{HoS}{
  short = HoS ,
  long  = Hours of Service,
  short-plural = s % Plural form of the short form
}

\DeclareAcronym{SoC}{
  short = SoC ,
  long  = State of Charge,
  short-plural = s % Plural form of the short form
}

\DeclareAcronym{MIP}{
  short = MIP ,
  long  = Mixed Integer Program,
  short-plural = s % Plural form of the short form
}

\DeclareAcronym{E-VSP}{
  short = E-VSP ,
  long  = Electric Vehicle Scheduling Problem,
  short-plural = s % Plural form of the short form
}

%% file: conference_101719.bbl
\begin{thebibliography}{00}

\bibitem{iea_2025}
IEA,
“Electric Vehicle Charging,” in \textit{Global EV Outlook 2025}, International Energy Agency, 2025.  
Available: \url{https://www.iea.org/reports/global-ev-outlook-2025/electric-vehicle-charging}. 

\bibitem{bjorklund_2025}
M.~Björklund, H.~Gillström, and F.~Stahre,
“Resilient electrified freight transport: Disruptions and mitigation strategies,”
\textit{Cleaner Logistics and Supply Chain}, vol.~14, 100211, March 2025,
doi: 10.1016/j.clscn.2025.100211.

% \bibitem{zanella_2025}
% A.~F.~Zanella, R.~H.~Palma Lima, M.~H.~Mulati, E.~V.~Cardoza Galdamez, and G.~C.~Lapasini Leal,
% “Vehicle routing and scheduling under hours of service regulations: A review,”
% \textit{Transportation Research Part A: Policy and Practice}, vol.~201, 104665, 2025, 
% doi: 10.1016/j.tra.2025.104665.

\bibitem{bai_2025}
T.~Bai, Y.~Li, A.~A.~Malikopoulos, K.~H.~Johansson, and J.~M{\aa}rtensson,
“Distributed Charging Coordination for Electric Trucks Under Limited Facilities and Travel Uncertainties,”
\textit{IEEE Transactions on Intelligent Transportation Systems}, vol.~26, no.~7, pp.~10278--10294, July 2025,
doi: 10.1109/TITS.2025.3550035.

\bibitem{barba_2025}
L.~Barba, S.~Ahmed, M.~Alamaniotis, and N.~Gatsis,
“Dynamic Modeling and Optimization of Long-Haul EV Charging Networks,”
in \textit{2025 IEEE Texas Power and Energy Conference (TPEC)}, College Station, TX, USA, 10–11 Feb. 2025, 
doi: 10.1109/TPEC63981.2025.10907255.

\bibitem{alhanahi_2024}
B.~Al-Hanahi, I.~Ahmad, D.~Habibi, P.~Pradhan, and M.~A.~S.~Masoum,
“A Charging Strategy for Large Commercial Electric Vehicle Fleets,”
\textit{IEEE Access}, vol.~12, pp.~46042--46058, 27 March 2024,
doi: 10.1109/ACCESS.2024.3382219.

\bibitem{alam_2023}
M.~R.~Alam and Z.~Guo,
“Co‑optimization of charging scheduling and platooning for long‑haul electric freight vehicles,”
\textit{Transportation Research Part C: Emerging Technologies}, vol.~147, 104009, February 2023,
doi: 10.1016/j.trc.2022.104009.

\bibitem{lee_2022}
J.-O.~Lee and Y.-S.~Kim,
“Novel battery degradation cost formulation for optimal scheduling of battery energy storage systems,”
\textit{International Journal of Electrical Power \& Energy Systems}, vol.~137, 107795, May 2022,
doi: 10.1016/j.ijepes.2021.107795.

\bibitem{gurobi} 
Gurobi Optimization, LLC.,
\textit{Gurobi Optimizer Reference Manual}, 2025.  
Available: \url{https://www.gurobi.com}.

\bibitem{scania_2025}
Scania Group,
“Battery Electric Trucks,” Scania Group website, 2025.  
Available: \url{https://www.scania.com/group/en/home/products-and-services/trucks/battery-electric-truck.html}.

\bibitem{weberlogistics_2025}
M.~Ma,
“How Retail Chargebacks Work and What You Can Do About Them,”
Weber Logistics California Logistics Blog, 2025.  
Available: \url{https://www.weberlogistics.com/blog/california-logistics-blog/how-retail-chargebacks-work-and-what-you-can-do-about-them}.



\end{thebibliography}
